\documentstyle[pra,aps,twocolumn,epsf]{revtex}

\begin{document}
\draft
\title{A Delayed Choice Quantum Eraser}
\author{Yoon-Ho Kim, R. Yu, S.P. Kulik\thanks{Permanent Address: Department of Physics,
Moscow State University, Moscow, Russia}, and Y.H. Shih}
\address{Department of Physics, University of Maryland, Baltimore County,\\
Baltimore, MD 21250}
\author{Marlan O. Scully}
\address{Department of Physics, Texas A \& M University, College Station, TX 77842\\
and Max-Planck Institut f\"{u}r Quantenoptik, M\"{u}nchen, Germany\\(submitted to PRL)}

\maketitle

\widetext

\begin{abstract}
This paper reports a ``delayed choice quantum eraser" experiment proposed by Scully and
Dr\"{u}hl in 1982. The experimental results demonstrated the possibility of
simultaneously observing both particle-like and wave-like behavior of a quantum via
quantum entanglement. The which-path or both-path information of a quantum can be erased
or marked by its entangled twin even after the registration of the quantum.
\end{abstract}


\narrowtext

Complementarity, perhaps the most basic principle of quantum mechanics, distinguishes the
world of quantum phenomena from the realm of classical physics. Quantum mechanically, one
can never expect to measure both precise position and momentum of a quantum at the same
time. It is prohibited. We say that the quantum observables ``position'' and ``momentum''
are ``complementary'' because the precise knowledge of the position (momentum) implies
that all possible outcomes of measuring the momentum (position) are equally probable. In
1927, Niels Bohr illustrated complementarity with ``wave-like'' and ``particle-like''
attributes of a quantum mechanical object \cite{Bohr}. Since then, complementarity is
often superficially identified with ``wave-particle duality of matter''. Over the years
the two-slit interference experiment has been emphasized as a good example of the
enforcement of complementarity. Feynman, discussing the two-slit experiment, noted that
this wave-particle dual behavior contains the basic mystery of quantum mechanics
\cite{Feynman}. The actual mechanisms that enforce complementarity vary from one
experimental situation to another. In the two-slit experiment, the common ``wisdom" is
that the position-momentum uncertainty relation $\delta x\delta p\geq \frac{\hbar }{2}$
makes it impossible to determine which slit the photon (or electron) passes through
without at the same time disturbing the photon (or electron) enough to destroy the
interference pattern. However, it has been proven \cite {Scully} that under certain
circumstances this common interpretation may not be true. In 1982, Scully and Dr\"{u}hl
found a way around this position-momentum uncertainty obstacle and proposed a quantum
eraser to obtain which-path or particle-like information without scattering or
\newpage \vspace*{30mm}
\noindent otherwise introducing large uncontrolled phase factors to disturb the
interference. To be sure the interference pattern disappears when which-path information
is obtained. But it reappears when we erase (quantum erasure) the which-path information
\cite{Scully,Wheeler}. Since 1982, quantum eraser behavior has been reported in several
experiments \cite{History}; however, the original scheme has not been fully demonstrated.

\begin{figure}[tbp]
\centerline{\epsfxsize=2.7in \epsffile{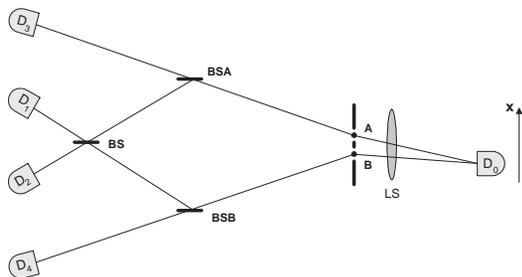}} \caption{A proposed quantum eraser
experiment. A pair of entangled photons is emitted from either atom A or atom B by atomic
cascade decay. ``Clicks" at $D_{3}$ or $D_{4}$ provide which-path information and
``clicks" at $D_{1}$ or $D_{2}$ erase the which-path information.}\label{fig:figure1}
\end{figure}

One proposed quantum eraser experiment very close to the 1982 proposal is illustrated in
Fig.\ref{fig:figure1}. Two atoms labeled by A and B are excited by a laser pulse. A pair
of entangled photons, photon 1 and photon 2, is then emitted from either atom A or atom B
by atomic cascade decay. Photon 1, propagating to the right, is registered by a photon
counting detector $D_{0}$, which can be scanned by a step motor along its $x$-axis for
the observation of interference fringes. Photon 2, propagating to the left, is injected
into a beamsplitter. If the pair is generated in atom A, photon 2 will follow the A path
meeting $BSA$ with 50\% chance of being reflected or transmitted. If the pair is
generated in atom B, photon 2 will follow the B path meeting $BSB$ with 50\% chance of
being reflected or transmitted. Under the 50\% chance of being transmitted by either
$BSA$ or $BSB$, photon 2 is detected by either detector $D_{3}$ or $D_{4}$. The
registration of $D_{3}$ or $D_{4}$ provides which-path information (path A or path B) of
photon 2 and in turn provides which-path information of photon 1 because of the
entanglement nature of the two-photon state of atomic cascade decay. Given a reflection
at either $BSA$ or $BSB$ photon 2 will continue to follow its A path or B path to meet
another 50-50 beamsplitter $BS$ and then be detected by either detector $D_{1}$ or
$D_{2}$, which are placed at the output ports of the beamsplitter $BS$. The triggering of
detectors $D_{1}$ or $D_{2}$ erases the which-path information. So that either the
absence of the interference or the restoration of the interference can be arranged via an
appropriately contrived photon correlation study. The experiment is designed in such a
way that $L_{0}$, the optical distance between atoms A, B and detector $D_{0}$, is much
shorter than $L_{i}$, which is the optical distance between atoms A, B and detectors
$D_{1}$, $D_{2}$, $D_{3}$, and $D_{4},$ respectively. So that $D_{0}$ will be triggered
much earlier by photon 1. After the registration of photon 1, we look at these
``delayed'' detection events of $D_{1}$, $D_{2}$, $D_{3}$, and $D_{4}$ which have
constant time delays, $\tau _{i}\simeq (L_{i}-L_{0})/c$, relative to the triggering time
of $D_{0}$. It is easy to see these ``joint detection'' events must have resulted from
the same photon pair. It was predicted that the ``joint detection'' counting rate
$R_{01}$ (joint detection rate between $D_{0}$ and $D_{1}$) and $R_{02}$ will show
interference pattern when detector $D_{0}$ is scanned along its $x$-axis. This reflects
the wave property (both-path) of photon 1. However, no interference will be observed in
the ``joint detection'' counting rate $R_{03}$ and $R_{04}$ when detector $D_{0}$ is
scanned along its $x$-axis. This is clearly expected because we now have indicated the
particle property (which-path) of photon 1. It is important to emphasize
that all four ``joint detection'' rates $R_{01}$, $R_{02}$, $R_{03}$, and $%
R_{04}$ are recorded at the same time during one scanning of $D_{0}$ along its $y$-axis.
That is, in the present experiment we ``see" both wave (interference) and which-path
(particle-like) with the same apparatus.

\begin{figure}[tbp]
\centerline{\epsfxsize=2.7in \epsffile{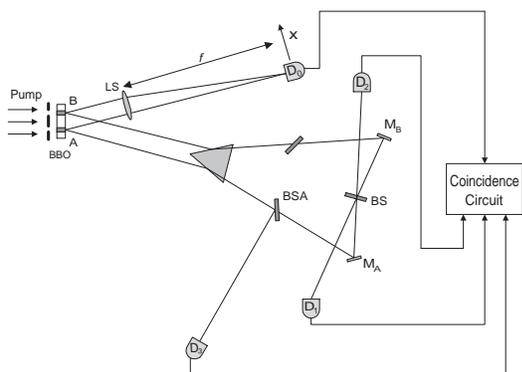}} \caption{Schematic of the
experimental setup. The pump laser beam of SPDC is divided by a double-slit and incident
onto a BBO crystal at two regions A and B. A pair of signal-idler photons is generated
either from A or B region. The detection time of the signal photon is $8ns$ earlier than
that of the idler.}\label{fig:figure2}
\end{figure}

We wish to report a realization of the above quantum eraser experiment. The schematic
diagram of the experimental setup is shown in Fig.\ref{fig:figure2}. Instead of atomic
cascade decay, spontaneous parametric down conversion (SPDC) is used to prepare the
entangled two-photon state. SPDC is a spontaneous nonlinear optical process from which a
pair of signal-idler photons is generated when a pump laser beam is incident onto a
nonlinear optical crystal \cite{SPDC}. In this experiment, the $351.1nm$ Argon ion pump
laser beam is divided by a double-slit and incident onto a type-II phase matching
\cite{typeII} nonlinear optical crystal BBO ($\beta-BaB_{2}O_{4}$) at two regions A and
B. A pair of $702.2nm$ orthogonally polarized signal-idler photon is generated either
from A or B region. The width of the SPDC region is about $0.3mm$ and the distance
between the center of A and B is about $0.7mm$. A Glen-Thompson prism is used to split
the orthogonally polarized signal and idler. The signal photon (photon 1, either from A
or B) passes a lens $LS$ to meet detector $D_{0}$, which is placed on the Fourier
transform plane (focal plane for collimated light beam) of the lens. The use of lens $LS$
is to achieve the ``far field'' condition, but still keep a short distance between the
slit and the detector $D_{0}$. Detector $D_{0}$ can be scanned along its $x$-axis by a
step motor. The idler photon (photon 2) is sent to an interferometer with equal-path
optical arms. The interferometer includes a prism $PS$, two 50-50 beamsplitters $BSA$,
$BSB$, two reflecting mirrors $M_{A}$, $M_{B}$, and a 50-50 beamsplitter $BS$. Detectors
$D_{1}$ and $D_{2}$ are placed at the two output ports of the $BS$, respectively, for
erasing the which-path information. The triggering of detectors $D_{3}$ and $D_{4}$
provide which-path information of the idler (photon 2) and in turn provide which-path
information of the signal (photon 1). The electronic output pulses of detectors $D_{1}$,
$D_{2}$, $D_{3}$, and $D_{4}$ are sent to coincidence circuits with the output pulse of
detector $D_{0}$,
respectively, for the counting of ``joint detection'' rates $R_{01}$, $%
R_{02} $, $R_{03}$, and $R_{04}$. In this experiment the optical delay ($%
L_{i}-L_{0} $) is chosen to be $\simeq 2.5m$, where $L_{0}$ is the
optical
distance between the output surface of $BBO$ and detector $D_{0}$, and $%
L_{i} $ is the optical distance between the output surface of the $BBO$ and detectors
$D_{1} $, $D_{2}$, $D_{3}$, and $D_{4}$, respectively. This means that any information
one can learn from photon 2 must be at least $8ns$ later
than what one has learned from the registration of photon 1. Compared to the $%
1ns$ response time of the detectors, $2.5m$ delay is good enough
for a ``delayed erasure''.

\begin{figure}[tbp]
\centerline{\epsfxsize=2.7in \epsffile{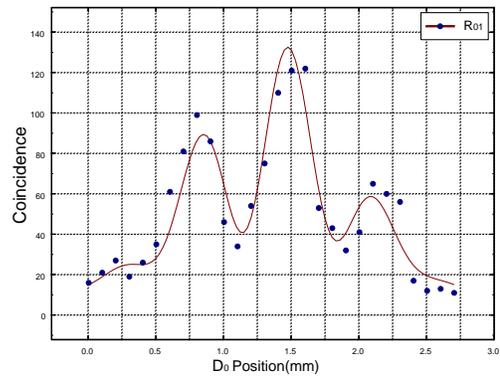}}
\caption{$R_{01}$ (``joint detection'' rate between detectors $D_{0}$ and $%
D_{1}$) against the $x$ coordinates of detector $D_{0}$. A standard Young's double-slit
interference pattern is observed.}\label{fig:figure3}
\end{figure}

\begin{figure}[tbp]
\centerline{\epsfxsize=2.7in \epsffile{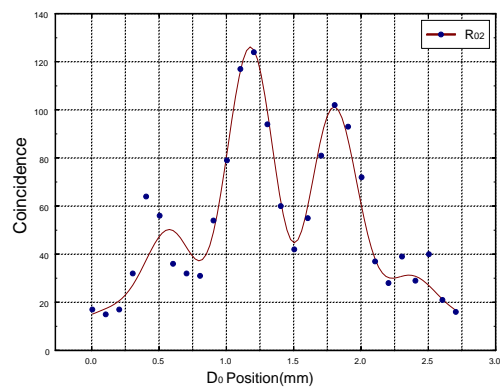}}
\caption{$R_{02}$ (``joint detection'' rate between detectors $D_{0}$ and $%
D_{2}$) Note, there is a $\pi$ phase shift compare to $R_{01}$ shown in
Fig.\ref{fig:figure3} }\label{fig:figure4}
\end{figure}

\begin{figure}[tbp]
\centerline{\epsfxsize=2.7in \epsffile{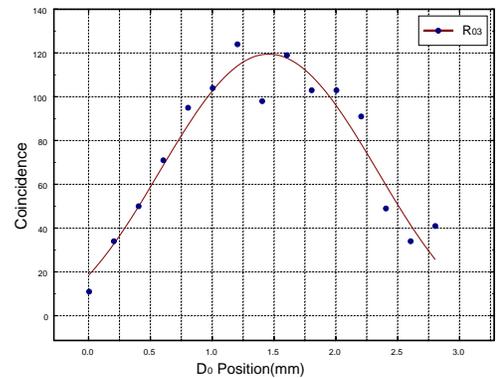}}
\caption{$R_{03}$ (``joint detection'' rate between detectors $D_{0}$ and $%
D_{3}$). An absence of interference is clearly demonstrated.}\label{fig:figure5}
\end{figure}

Figs.\ref{fig:figure3}, \ref{fig:figure4}, and \ref{fig:figure5} report the experimental
results, which are all consistent with prediction. Figs.\ref{fig:figure3} and
\ref{fig:figure4} show the ``joint detection'' rates $R_{01}$ and $R_{02}$ against the
$x$ coordinates of detector $D_{0}$. It is clear we have observed the standard Young's
double-slit interference pattern. However, there is a $\pi $ phase shift between the two
interference fringes. The $\pi$ phase shift
is explained as follows. Fig.\ref{fig:figure5} reports a typical $%
R_{03}$ ($R_{04}$), ``joint detection'' counting rate between $D_{0}$ and ``which-path" $%
D_{3}$ ($D_{4}$), against the $x$ coordinates of detector $D_{0}$. An absence of
interference is clearly demonstrated. There is no significant difference between the
curves of $R_{03}$ and $R_{04}$ except the small shift of the center.

To explain the experimental results, a standard quantum mechanical calculation is
presented in the following. The ``joint detection'' counting rate, $R_{0i}$, of detector
$D_{0}$ and detector $D_{j},$ on the time interval $T$, is given by the Glauber formula
\cite{Glauber}:
\begin{eqnarray}
R_{0j}&\propto& \frac{1}{T}\int_{0}^{T}\int_{0}^{T}dT_{0}dT_{j}\langle \Psi |
E_{0}^{(-)}E_{j}^{(-)}E_{j}^{(+)}E_{0}^{(+)}| \Psi \rangle \nonumber \\
&=&\frac{1}{T}\int_{0}^{T}\int_{0}^{T}dT_{0}dT_{j}| \langle 0| E_{j}^{(+)}E_{0}^{(+)}|
\Psi \rangle | ^{2} , \label{coin}
\end{eqnarray}
where $T_{0}$ is the detection time of $D_{0}$, $T_{j}$ is the detection time of $D_{j}($
$j=1,2,3,4$) and $E_{0,j}^{(\pm )}$ are positive and negative-frequency components of the
field at detectors $D_{0}$ and $D_{j}$, respectively. $| \Psi \rangle$ is the entangled
state of SPDC,

\begin{equation}
| \Psi \rangle =\sum_{s,i} C({\bf k}_{s}, {\bf k}_{i}) \ a_{s}^{\dagger }(\omega ({\bf
k}_{s}))\ a_{i}^{\dagger }(\omega ({\bf k}_{i}))| 0\rangle \label{state} ,
\end{equation}
where $C({\bf k}_{s}, {\bf k}_{i})=\delta( \omega_{s}+\omega_{i}-\omega_{p}) \delta( {\bf
k}_{s}+{\bf k}_{i}-{\bf k}_{p})$, for the SPDC in which $\omega_{j}$ and ${\bf k_{j}}$
$(j = s, i, p)$ are the frequency and wavevectors of the signal ($s$), idler ($i$), and
pump ($p$), respectively, $\omega_{p}$ and ${\bf k}_{p}$ can be considered as constants,
a single mode laser line is used for pump and $a_{s}^{\dagger }$ and $a_{i}^{\dagger }$
are creation operators for signal and idler photons, respectively. For the case of two
scattering atoms, see ref. \cite{Scully}, and in the case of cascade radiation, see ref.
\cite{Scully2}, $C({\bf k}_{s}, {\bf k}_{i})$ has a similar structure but without the
momentum delta function. The $\delta$ functions in eq.(\ref{state}) are the results of
approximations for an infinite size SPDC crystal and for infinite interaction time. We
introduce the two-dimensional function $\Psi (t_{0}, t_{j})$ as in eq.(\ref{coin}),
\begin{equation}
\Psi(t_{0}, t_{j})\equiv \langle 0| E_{j}^{(+)}E_{0}^{(+)}| \Psi \rangle .
\label{wavefunction}
\end{equation}
$\Psi(t_0,t_j)$ is the joint count probability amplitude (``wavefunction" for short),
where $t_{0}\equiv T_{0}-L_{0}/c,$ $%
t_{j}\equiv T_{j}-L_{j}/c, j=1,2,3,4,$ $L_{0}$ ($L_{j}$) is the optical distance between
the output point on the BBO crystal and $D_{0}$ ($D_{j}$). It is straightforward to see
that the four ``wavefunctions'' $\Psi (t_{0}, t_{j})$, correspond to four different
``joint detection'' measurements, having the following different forms:
\begin{eqnarray}
&\Psi(t_{0}, t_{1}) = A(t_{0},t_{1}^{A})+A(t_{0},t_{1}^{B}),& \nonumber \\ &\Psi (t_{0},
t_{2}) =A(t_{0},t_{2}^{A})-A(t_{0},t_{2}^{B}) ,& \label{R12}  \\
&\Psi(t_{0},t_{3})=A(t_{0},t_{3}^{A}),\quad \Psi (t_{0},t_{4})=A(t_{0},t_{4}^{B}) ,&
\label{R34}
\end{eqnarray}
where as in Fig.\ref{fig:figure1} the upper index of $t$ (A or B) labels the scattering
crystal (A or B region) and the lower index of $t$ indicates different detectors. The
different sign between the two amplitudes $\Psi (t_{0}, t_{1})$ and $\Psi (t_{0}, t_{2})$
is caused by the transmission-reflection unitary transformation of the beamsplitter $BS$,
see Fig.\ref{fig:figure1} and Fig.\ref{fig:figure2}. It is also straightforward to
calculate each of the $A(t_{i}, t_{j})$ \cite{function}. To simplify the calculations, we
consider the longitudinal integral only and write the two-photon state in terms of the
integral of $k_{e}$ and $k_{o}$:
\begin{eqnarray}
| \Psi \rangle =A_{0}^{^{\prime}}\int dk_{e}\int dk_{o}\
\delta(\omega_{e}+\omega_{o}-\omega_{p}) \times \nonumber \\ \Phi(\Delta_{k}L)
a_{k_{e}}^{\dagger}a_{k_{o}}^{\dagger}|0\rangle ,  \label{B-2}
\end{eqnarray}
where a type-II phase matching crystal with finite length of $L$ is assumed.
$\Phi(\Delta_{k}L)$ is a sinc-like function,
$\Phi(\Delta_{k}L)=(e^{i(\Delta_{k}L)}-1)/i(\Delta_{k}L)$. Using eqs.
(\ref{wavefunction}) and (\ref{B-2}) we find,
\begin{eqnarray}
A(t_{i}, t_{j})=A_{0}\int dk_{e}\int dk_{o} \delta(\omega_{e}+\omega_{o}-\omega_{p})
\times \nonumber \\
\Phi(\Delta_{k}L) f_{i}(\omega_{e})f_{j}(\omega_{o})
e^{-i(\omega_{e}t_{1}^{e}+\omega_{o}t_{2}^{o})} , \label{B-5}
\end{eqnarray}
where $f_{i,j}(\omega ),$ is the spectral transmission function of an assumed filter
placed in front of the $k_{th}$ detector and is assumed Gaussian to simplify the
calculation. To complete the integral, we define $\omega_{e}=\Omega_{e}+\nu$ and
$\omega_{o}=\Omega_{o}-\nu$, where $\Omega_{e}$ and $\Omega_{o}$ are the center
frequencies of the SPDC, $\Omega_{e}+\Omega_{o}=\Omega_{p}$ and $\nu$ is a small tuning
frequency, so that $\omega_{e}+\omega_{o}=\Omega_{p}$ still holds. Consequently, we can
expand $k_{e}$ and $k_{o}$ around $K_{e}(\Omega_{e})$ and $K_{o}(\Omega_{o})$ to first
order in $\nu$:
\begin{eqnarray}
k_{e} &=& K_{e}+\nu \left.\frac{d\omega_{e}}{dk_{e}}\right|_{\Omega_{e}} = K_{e}+%
\frac{\nu}{u_{e}} , \nonumber \\ k_{o} &=& K_{o}-\nu
\left.\frac{d\omega_{o}}{dk_{o}}\right|_{\Omega_{o}} = K_{o}-\frac{\nu}{u_{o}} ,
\label{B-6}
\end{eqnarray}
where $u_{e}$ and $u_{o}$ are recognized as the group velocities of the e-ray and o-ray
at frequencies $\Omega_{e}$ and $\Omega_{o}$, respectively. Completing the integral, the
biphoton wavepacket of type-II SPDC is thus:
\begin{equation}
A(t_{i}, t_{j})=A_{0} \Pi(t_{i}-t_{j}) e^{-i\Omega_{i}t_{i}} e^{-i\Omega_{j}t_{j}} ,
\label{B-14}
\end{equation}
where we have dropped the $e,o$ indices. The shape of $\Pi(t_{1}-t_{2})$ is determined by
the bandwidth of the spectral filters and the parameter $DL$ of the SPDC crystal, where
$D\equiv 1/u_{o}-1/u_{e}$. If the filters are removed or have large enough bandwidth, we
have a rectangular pulse function $\Pi(t_{1}-t_{2})$.
\[
\Pi(t_0-t_j) =
    \left\{
        \begin{array}{ll}
            1 & {\rm if~} 0\leq t_{0}-t_{j}\leq DL , \\
            0 & {\rm otherwise}.
        \end{array}
    \right.
\]
It is easy to find that the two amplitudes in $\Psi(t_{0}, t_{1})$ and $\Psi(t_{0},
t_{2})$ are indistinguishable (overlap in both $t_{0}-t_{j}$ and $t_{0}+t_{j}$),
respectively, so that interference is expected in both the coincidence counting rates,
$R_{01}$ and $R_{02}$; however, with a $\pi$ phase shift due to the different sign,
\[
R_{01}\propto \cos ^{2}(x\pi d/\lambda f),\quad {\rm and} \quad R_{02}\propto \sin
^{2}(x\pi d/\lambda f).
\]
If we consider ``slit'' A and B both have finite width (not infinitely narrow), an
integral is necessary to sum all possible amplitudes along slit
A and slit B. We will have a standard interference-diffraction pattern for $%
R_{01}$ and $R_{02}$,
\begin{eqnarray}
R_{01} &\propto&{\rm sinc}^{2}(x\pi a/\lambda f)\cos^{2}(x\pi d/\lambda f) , \nonumber \\
R_{02} &\propto&{\rm sinc}^{2}(x\pi a/\lambda f)\sin^{2}(x\pi d/\lambda f) ,\label{R0102}
\end{eqnarray}
where $a$ is the width of the slit A and B (equal width), $d$ is the distance between the
center of slit A and B, $\lambda =\lambda_{s}=\lambda_{i}$ is the wavelength of the
signal and idler, and $f$ is the focal length of lens $LS$. We have also applied the
``far field approximation'' for the signal and equal optical distance of the
interferometer for the idler. After considering the finite size of the detectors and the
divergence of the pump beam for further integrals, the interference visibility is reduced
to the level close to the observation.

For the ``joint detection'' $R_{03}$ and $R_{04}$, it is seen that the ``wavefunction''
in eq.(\ref{R34}) (which clearly provides ``which-path'' information) has only one
amplitude and no interference is expected.

In conclusion, we have realized a quantum eraser experiment of the type proposed in ref.
\cite{Scully}. The experimental results demonstrate the possibility of observing both
particle-like and wave-like behavior of a light quantum via quantum mechanical
entanglement. The which-path or both-path information of a quantum can be erased or
marked by its entangled twin even after the registration of the quantum.

This work was supported, in part, by the U.S. Office of Naval Research, the Army Research
Office - the National Security Agency, the National Science Foundation, and the Welch
Foundation. MOS wishes to thank Roland Hagen for helpful and stimulating discussions.

\begin{references}
\bibitem{typeII}  In type-I SPDC, signal and idler are both ordinary rays of
the crystal; however, in type-II SPDC the
signal and idler are orthogonal polarized, i.e., one is ordinary
ray and the other is extraordinary ray of the crystal.
\end{references}
\end{document}